\documentclass[%
 amsmath,amssymb,
 reprint, nofootinbib
]{revtex4-1}
 
\usepackage{graphicx}
\usepackage{dcolumn}
\usepackage{bm}

\usepackage[utf8]{inputenc}
\usepackage[T1]{fontenc}
\usepackage{mathptmx}
\usepackage{etoolbox}
\usepackage{amsmath}

\makeatletter
\def\@email#1#2{%
 \endgroup
 \patchcmd{\titleblock@produce}
  {\frontmatter@RRAPformat}
  {\frontmatter@RRAPformat{\produce@RRAP{#1\href{mailto:#2}{#2}}}\frontmatter@RRAPformat}
  {}{}
}%
\makeatother
\begin{document}

\title{Quantum Effects in Ion Transport: A Thermodynamic Resource Theory Approach}

\author{Amin Mohammadi}

\author{Afshin Shafiee}

\email{shafiee@sharif.edu}
\email{amin.mohammadi@ch.sharif.edu}

\affiliation{ 
Research Group on Foundations of Quantum Theory and Information, Department of Chemistry,
Sharif University of Technology P.O.Box 11365-9516, Tehran, Iran}

\begin{abstract}
In recent years, understanding thermodynamics in the quantum regime has garnered significant attention, driven by advances in nanoscale physics and experimental techniques. In parallel, growing evidence supports the importance of quantum effects in various biological processes, making them increasingly relevant to quantum thermodynamics.
In this study, we apply resource theory formulations of thermodynamics to investigate the role of quantum properties in ion transport across cell membranes. Within this framework, quantum properties are treated as resources under generalized thermodynamic constraints in the quantum regime. Specifically, our findings reveal that non-Markovianity, which reflects memory effects in ion transport dynamics, is a key quantum resource that enhances the yield and efficiency of the ion transport process. In contrast, quantum coherence, manifested as the superposition of energy states in ion-transport proteins, reduces these metrics but plays a crucial role in distinguishing between ion channels and ion pumps— two distinct types of ion-transport proteins in cell membranes. Finally, we demonstrate that introducing an additional coherent system allows coherence to facilitate the transformation of an ion pump into an ion channel.
\end{abstract}

\maketitle

\section{\label{sec:level1}Introduction}
The selective conduction of specific ions into and out of cells is at the heart of many biological processes. Some relevant examples include cell volume regulation, secretion, and electrical signal generation in nerve and muscle cells \cite{hille2001ion}.
Ion channels and ion pumps are two different classes of ion-transport proteins that are used by the cell to transport ions across the membranes \cite{gouaux2005principles,gadsby2009ion}. Ion channels transport ions in line with their natural tendency to flow from regions of higher to lower electrochemical potential while ion pumps do the opposite task, building up electrochemical gradients, at the expenses of the energy typically supplied by the hydrolysis of ATP \cite{hille1978ionic,aidley1996ion,kuhlbrandt2004biology,moller1996structural,gouaux2005principles}. In this respect, ion transport within channels can be described as a passive, thermodynamically downhill process, as opposed to the active, thermodynamically uphill process of ion transport in pumps \cite{gouaux2005principles,gadsby2009ion}. 

A particular consequence of these distinct functions is the higher speed of ion transport through channels that often are several magnitudes of order faster than ion transport through pumps \cite{gouaux2005principles,gadsby2009ion}. In the potassium channels, for instance, K+ ions flow at a rate range of $10^6-10^8\ s^{-1}$ while this range in sodium/potassium pumps is about $10^2\ s^{-1}$ \cite{roux2017ion,mackinnon2004potassium}. An important property, however, remains common in both channels and pumps: they are very selective for specific ions \cite{gouaux2005principles,gadsby2009ion}.
Ion channels are also equipped with at least one gate that can be opened and closed by conformational changes, regulating the flow of ions through the channels, on a time scale similar to the rate range in which ions are pumping \cite{gadsby2009ion}.

Such observations have led some scientists to suggest that channels and pumps may be thought of as fundamentally much alike \cite{gadsby2009ion,patlak1957contributions,vidaver1966inhibition,jardetzky1966simple,lauger1979channel,gadsby1993extracellular}.
What forms the basis for the similarity is looking at a pump as a channel with two gates and what makes the difference is restricting two gates from opening at once \cite{gadsby2009ion}. 
Hence, in this view, the principle for the protein to function as a pump lies in a fail-safe mechanism that ensures the gates do not open at the same time, but rather open and close alternately to limit the speed of ion movement through pumps.
If such a mechanism does not work, the protein will act as an ion channel with two separate open gates \cite{gadsby2009ion}.

Experimental evidence in favor of the above proposal was provided by Artigas and Gadsby \cite{artigas2003na+}. They added palytoxin, a deadly biotoxin, to a single sodium/potassium pump and detected an immediate increase in the electrical current from zero to nearly a trillionth of an ampere.
These recorded currents align respectively with the slow movements of ions through pumps, producing non-measurable signals, and rapid movements of ions within channels. 
In other words, the experiment demonstrated that palytoxin interfered with the pump's gating mechanism allowing two gates to be opened simultaneously.

On the other hand, ion channels are the subject of a more recent avenue of research seeking the possible existence and functional relevance of quantum phenomena such as quantum coherence \cite{mohseni2014quantum,ganim2011vibrational,salari2017quantum,vaziri2010quantum,salari2015quantum,seifi2022quantum,seifi2024mimicking}.
These possibilities align with prior studies that explore transient quantum effects in other biological processes like photosynthesis \cite{romero2014quantum,lloyd2011quantum,ghasemi2020investigation}.
For example, authors in \cite{vaziri2010quantum} model the ion transport in the potassium channel with N qubits interacting in a tight-binding-type Hamiltonian, representing the different sites in which the ions may be located during the transport.
They then exposed the channel to the constant and time-dependent potentials
and solved the dynamics of the model in the Lindblad formalism, including various types of environmental effects.
They demonstrated that, for certain choices of model parameters, the channel exhibits quantum coherence which suppresses its conductivity.
In \cite{seifi2022quantum}, it was also shown that quantum coherence is present in the dynamics of the potassium channel modeled as a spin-1 system interacting with the environment in the Lindblad formalism. 
It was observed, however, that the high transfer efficiency in the channel not only does not contradict the presence of quantum coherence in this model, but a high throughput rate of ions is necessary for the channel to remain coherent.   

Therefore, the results obtained in such studies are greatly constrained by the particular models and particular dynamics, resulting even in potentially conflicting outcomes. 
In fact, these models extremely simplify the dynamics of ion channels that are inherently highly complex due to the presence of multiple interacting degrees of freedom.
Such complexity also poses obstacles for the experimental and computational tools in yielding conclusive outcomes \cite{ganim2011vibrational, roux2004computational}. 
So it would be beneficial to perform such a study using methods that are minimal on the details of dynamics.
Such an abstract view of dynamics is common in many quantum information tools such as resource theories, rendering them versatile for various applications \cite{chitambar2019quantum,streltsov2017colloquium,nielsen2010quantum}.
In particular,
resource theories have developed approaches for extending thermodynamics to quantum regime known as resource theories of athermality and asymmetry \cite{horodecki2013fundamental,brandao2013resource,brandao2015second,cwiklinski2015limitations,lostaglio2015quantum,lostaglio2015description,lostaglio2019introductory}.
Relative to the
energy eigenbasis, the former generalizes the thermodynamic laws constraining the transformations on the population of the quantum states while the latter accounts for the thermodynamic processing of quantum coherence. 
Resource theory of athermality has been recently applied, for example, to study the photoisomerization probability of a molecular switch \cite{yunger2020fundamental,spaventa2022capacity}.   

In the present study, we model the ion transport across the cell membrane within the thermodynamic resource theories.
We evaluate the performance of the ion transport process by considering two quantities: first, we define the transport yield, which indicates the population of ions at the end of the transport process; second, we examine the entropy production, which represents the fundamental loss experienced during the transport process and serves as a measure of its efficiency.
The transport yield is assessed using the resource theory of athermality. While quantum coherence does not directly affect the transport yield, it contributes to entropy production, which can be quantified using the resource theory of asymmetry.
In addition, we investigate how non-Markovian behavior and resulting memory effects in ion transport dynamics influence both transport yield and efficiency.
We propose a fail-safe mechanism that differentiates between ion channels and ion pumps.
In this mechanism, the transport process in ion pumps is limited to operate at the maximum yield and and is prevented from exhibiting quantum coherence.
However, the second law of thermodynamics prevents this process from meeting these criteria, indicating that ion pumps cannot transport ions unaided.
Given that quantum coherence does not put any additional constraints, the thermodynamic constraint on ion transport inside a pump can be lifted by introducing battery states that supply the enough energy.
Finally, we discuss how an external coherent source can bypass the fail-safe mechanism, converting a pump into a channel.

In the remainder of the paper, we first briefly review the concept of thermodynamics resource theories in the subsequent section before presenting our model in section III and discussing our results in section IV. We end the paper with the conclusions in section V.
\section{thermodynamics resource theories }
\subsection{\label{sec:level2}Free operations and free states }
Given some physical constraints, a resource theory is defined by a set of allowed operations known as free operations and a set of states that are only achievable through free operations known as free states. Any states beyond the free set are considered a resource, whose creation requires the costly implementation of operations outside the set of free operations.
Resource theories have been successfully utilized in quantum physics to model a variety of operational tasks, accounting for the advantages that can be brought about by exploiting the relevant quantum properties.
In this respect, thermodynamics has been recast in a resource theoretic framework extending its laws to cover the non-equilibrium dynamics of arbitrary quantum systems \cite{horodecki2013fundamental,brandao2013resource,brandao2015second,cwiklinski2015limitations,lostaglio2015quantum,lostaglio2015description,lostaglio2019introductory}. 
Within this framework, the state transformation of a quantum system in contact with a bath at a fixed temperature is restricted to conserve energy, a constraint imposed by the first law of thermodynamics. 
The set of thermodynamically allowed operations can be then considered as following three operations:(1) preparing the bath at a thermal equilibrium state at a fixed temperature T and with an arbitrary Hamiltonian $H_B$, (2) performing an arbitrary energy-conserving unitary on the system and bath, (3) discarding arbitrary subsystems. 
These free operations can be combined to represent a quantum channel, known as thermal operation, mapping the state of the system $\rho_S$ with the Hamiltonian $H_S= \sum_x E_x |x>_S<x|$ as follows:
\begin{equation}
  \mathcal{T}(\rho_S)= Tr_B[U (\rho_S \otimes \gamma_B) U^\dagger]
\end{equation}
where $\gamma_B = e^{-\beta H_B}/Z_B$ is the Gibbs state denoting the bath in thermal equilibrium, $Z_B=Tr[e^{-\beta H_B}]$ is the partition function for bath, and $U$ is any global unitary satisfying $[U, H_S+H_B]=0$. Thermal operations satisfy two important properties.
First, they do not disturb the system in the thermal equilibrium, i.e., they are Gibbs-preserving operations:
\begin{equation}
  \mathcal{T}(\gamma_S)=\gamma_S
  \label{eq:2}
\end{equation}
where $\gamma_S$ is Gibbs state of system.
Second, they are symmetric under time translations of system:
\begin{equation}
  \mathcal{T}(e^{-iH_S t} (\rho_S) e^{iH_S t})= e^{-iH_S t}(\mathcal{T}(\rho_S))e^{iH_S t}
  \label{eq:3}.
\end{equation}
Noting that no work can be extracted from the system in a Gibbs state and incoherent states are those diagonal in energy eigenbasis, $[\rho_S, H_S]=0$,  Eqs.~(\ref{eq:2}) and (\ref{eq:3}) imply that no work and no energetic coherence can be produced for free in the thermodynamic processing of a quantum system.

Furthermore, as a result of time translation symmetry, the evolution of the diagonal elements of the density matrix $\rho_S$ (populations) is decoupled from the evolution of off-diagonal ones (coherences) for thermal operations. 
More precisely, a thermal operation maps the diagonal matrix elements of a general quantum state, $\rho_S=\sum_{x.y} \rho_{x,y} |x>_S<y|$, as:
\begin{equation}
  \mathcal{T}(|x>_S<x|)= \sum_{x^{'}} G_{x^{'}|x} |x^{'}>_S<x^{'}|
  \label{eq:4}
\end{equation}
and provide a upper bound for the evolution of non-diagonal elements:
\begin{equation}
  \mathcal{T}(|x>_S<y|) = \sum_{x^{'},y{'}}^{\omega_{xy}} {C_{y{'}|y}^{x{'}|x}}  |x^{'}>_S<y^{'}|, \hspace{0.5cm}|C_{y{'}|y}^{x{'}|x}| \leq \sqrt{G_{x{'}|x} G_{y^{'}|y}}
  \label{eq:5}.
\end{equation}
The coefficients $G_{x^{'}|x}$ are elements of a Gibbs stochastic matrix G, i.e., stochastic matrices ($G_{y|x} \geq 0$, $\sum_y G_{y|x}=1$) that preserve Gibbs states ($G \gamma = \gamma$, $\gamma_i=e^{-\beta E_i^S}/Z_S$).
The symbol $\sum^{\omega_{xy}}$ indicates the sum over indices $x$,$y$ are taken for each mode of coherence $\omega_{xy}= E_x^S-E_y^S$ \cite{lostaglio2015quantum,lostaglio2019introductory} and $C_{y{'}|y}^{x{'}|x}$ are coefficients by which the non-diagonal elements are damped during evolution.

Therefore, considering the thermal operations as free operations, a resource theory of thermodynamics can be segmented into two parts: athermality and asymmetry, with each being accountable for thermodynamic laws that constrain the dynamics of population and quantum coherence.

In the resource theory of athermality, Gibbs states construct the set of free states with the remaining diagonal states comprising the set of resources. Here, thermal operations are equivalent to Gibbs-stochastic matrices as shown in Eq.~(\ref{eq:4}).
In other words, given two diagonal states $\rho$ and $\sigma$ with corresponding population vectors $x$ and $y$, thermodynamics constrains the possibility of a transition from $\rho$ to $\sigma$, to the existence of a Gibbs-stochastic matrix such that $x= G y$. 
This problem can be addressed using the thermomajorization criterion \cite{horodecki2013fundamental}.
According to it, one associates with the population vectors $x$ and $y$  the thermomajorization curves $\emph{T}(x)$ and $\emph{T}(y)$ (see, e.g., Supplementary Note 2 in \cite{horodecki2013fundamental}) and say that $x$ thermomajorize $y$, denoted $x \succ_g y$, if and only if the curve $\emph{T}(x)$ is everywhere on or above the curve $\emph{T}(y)$. Then, there exist a thermal operation mapping $\rho$ to $\sigma$ if and only if $x \succ_g y$.
 
Conversely, in the resource theory of asymmetry, all diagonal states are considered free, and resources are identified as non-diagonal (coherent) states, lacking symmetry under time translations.
Here, thermomajorization remains a necessary criterion for state transitions limiting the evolution of diagonal elements of quantum states but becomes insufficient due to the presence of non-diagonal elements.    
Indeed, the time translation symmetry of thermal operations (Eq.~(\ref{eq:3})) places additional constraints on the evolution of non-diagonal elements of quantum states that need to be satisfied independently of thermomajorization.
These constraints enforce the quantum coherence to get degraded during evolution under thermal operations.

\subsection{\label{sec:level2}Resource monotones }
A key feature of resource theories is the idea of resource monotones, i.e. non-negative real numbers whose values can not increase under application of any free operation. Resource monotones quantify how far a non-free state is from being free and then provide a measure of the resourcefulness of quantum states. 
As shown, thermodynamics can be viewed as composed of two independent resources, each needing to be quantified individually. Two parallel sets of monotones have been demonstrated for this purpose: first, a continuum of generalized free energies given by:
\begin{equation}
  F_\alpha (x)= -\emph{K}T log Z_S + \emph{K}T \emph{S}_\alpha(x||\gamma)
  \label{eq:6}.
\end{equation}
for any $\alpha \geq 0$, to quantify athermality, i.e., how far a population vector $x$ (or diagonal state $\rho_S$) is from being Gibbsian \cite{brandao2015second}
and a continuum of generalized asymmetry monotones given by:
\begin{equation}
  A_\alpha (\rho_S)= \widetilde{\emph{S}}_\alpha(\rho_S||D(\rho_S))
  \label{eq:7}.
\end{equation}
for any $\alpha \geq 0$, to quantify asymmetry, i.e., how far a quantum state $\rho_S$ is from being incoherent in energy eigenbasis \cite{lostaglio2015description}, which $D(\rho_S)$ denotes the
state obtained from $\rho_S$ by removing all off-diagonal elements.
$\emph{S}_\alpha$ in Eq.~(\ref{eq:6}) are $\alpha$-Renyi divergences \cite{renyi1961measures}, defined by
\begin{equation}
\emph{S}_\alpha(x||y)= \frac{sgn(\alpha)}{\alpha - 1} \log \sum_{i} x_i^\alpha y_i^{1 - \alpha} 
\end{equation}
and $\widetilde{\emph{S}}_\alpha$ in Eq.~(\ref{eq:7}) are the quantum Renyi divergences \cite{muller2013quantum,wilde2014strong,mosonyi2015quantum}, defined by
\begin{equation}
\widetilde{S}_\alpha(\rho_S \|\sigma_S) =
\begin{cases} 
\frac{1}{\alpha - 1} \log \operatorname{Tr} \left[ \rho_S^\alpha \sigma_S^{1-\alpha} \right], & \alpha \in (0,1), \\
\frac{1}{\alpha - 1} \log \operatorname{Tr} \left[ (\sigma_S^{\frac{1-\alpha}{2\alpha}} \rho_S \, \sigma_S^{\frac{1-\alpha}{2\alpha}} )^\alpha \right], & \alpha > 1. 
\end{cases}
\end{equation}

Therefore, for any state transition under thermal operations, one necessarily has $\Delta F_\alpha\leq0$ and $\Delta A_\alpha\leq0$ for all $\alpha \geq 0$. Former can also be sufficient in the case of catalytic thermal operations, which allow the use of auxiliary systems that are returned to their initial state at the end \cite{brandao2015second,lostaglio2019introductory}.

\section{Resource theory model for the ion transport }
\begin{figure*} [hbt!]
\includegraphics[width=6in, height=4in]{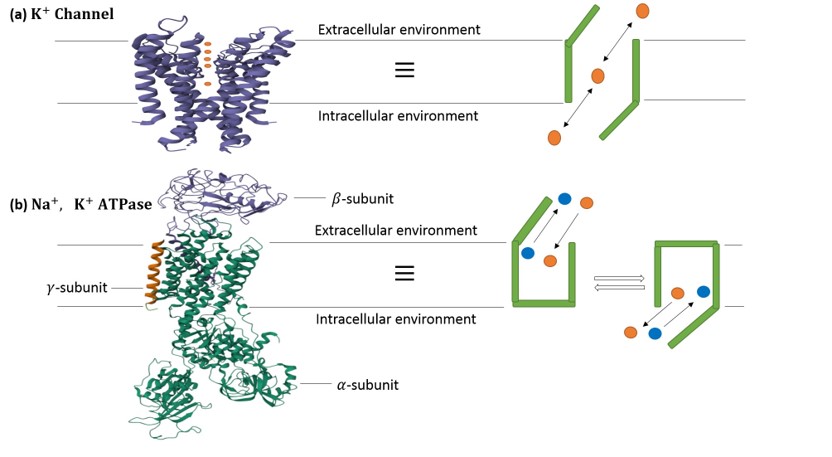}
\caption{\label{fig:1} Effective representations of an ion-transport protein illustrate it as a membrane-spanning pore that regulates the movement of ions (colored spheres) via two gates, considered as a two-level quantum system:
(a) an open ion channel featuring two simultaneously open gates and (b) an open ion pump characterized by two gates that alternate between open and closed situations to export and import two different kinds of ions.
Unlike ion channels, which transport only one particular type of ion, most ion pumps transport different kinds of ions in opposite directions across cell membranes.
The left sides in (a) and (b) correspond to the real structures of the KcsA K ion channel and Na, K ATPase pump with PDB codes 1K4C and 3B8e, respectively.}
\end{figure*}
We now express our model to discuss the ion transport across cell membranes within thermodynamics resource theories.
We consider the flow of ions to be controlled by two gates located along the translocation pathway within membrane proteins, as illustrated in Fig.~(\ref{fig:1}) .
Upon the first gate opening, ions from the extracellular (intracellular) environment flow through the protein, whereas the second gate opening permits ions to enter the intracellular (extracellular) environment.
This pivotal role of gates suggests that the membrane protein can be effectively represented as a two-level system, with each level corresponding to the opening of a gate. 
We further assume the biological environments to be a thermal bath at homogeneous temperature $T$ and that all interactions between the system and bath are due to an energy-preserving unitary $U$.
While these abstractions avoid reliance on specific details of real biological environments, such as protein-water or lipid-protein interactions, they capture the essential thermodynamic features needed to explore fundamental principles underlying quantum effects in ion transport.
Importantly, the assumption of thermality aligns with resource theory of thermodynamics, where no further constraints are placed on the environment beyond its thermal nature \cite{lostaglio2019introductory}.
In consequence, we can now describe the dynamics of ion transport in the membrane protein as a mapping between protein states via a thermal operation, i.e., $\rho^f=\mathcal{T}(\rho^i)$, where $\rho^i$ and $\rho^f$ denote the protein states before and after the transport process.
The two-level system representation of protein also allows for a complete characterization of the dynamics of the quantum coherence in our model, as the bound of Eq.~(\ref{eq:5}) is only achievable in this particular case. For two-level systems, the most general form of Gibbs-stochastic matrix acting on the populations can be written as
\begin{equation}
G(E,\beta,\lambda)= (1-\lambda) \begin{bmatrix}
1 & 0\\ 
0 & 1
\end{bmatrix}
+
\lambda
\begin{bmatrix}
1 - e^{-\beta E} & 1\\ 
e^{-\beta E} & 0
\end{bmatrix}
\label{eq:8}
\end{equation}
where $E$ is the system energy gap and parameter $\lambda$ is a convex parameter $\lambda\in [0,1]$, describing the strength of interaction with bath \cite{cwiklinski2015limitations,lostaglio2019introductory}.
The state transformation under thermal operations 
can be then written as $\rho^f=\sum_\omega K_\omega (\rho^i)K^\dagger_\omega$ with the following Kraus operators 
\begin{multline}
 k_0= \sqrt{G_{0|0}} |0><0|+ \sqrt{G_{1|1}} |1><1|,\\     k_1=\sqrt{G_{1|0}} |1><0|,\hspace{1.2cm}     k_{-1}=\sqrt{G_{0|0}} |0><1|.
\end{multline}

We define the transport yield as the population at the second gate in the final state, denoted by $Y=<1|\rho^f|1>$. This metric represents the proportion of ions that successfully traverse both gates to reach the cellular environment, providing essential insights into the functional performance of membrane proteins.
However, it does not account for the thermodynamic costs associated with the process.
Since ion transport occurs in a non-equilibrium setting, its thermodynamic efficiency is inherently constrained by the entropy generated through the irreversible dissipation of energy into the environment \cite{lebon2008understanding}. As a result, the ion transport performance can also be complementarily characterized by its thermodynamic efficiency, which is quantified through entropy production as follows:
\begin{equation}
 \Sigma= S(\rho^i||\gamma)- S(\rho^f||\gamma)
\end{equation}
where $S(\rho_S||\gamma_S)=Tr[\rho_S(log(\rho_S)-log(\gamma_S))]$ is the relative entropy between the non-equilibrium and equilibrium states \cite{landi2021irreversible}. 
Maximum efficiency is achieved when entropy production is zero ($\Sigma=0$), corresponding to a fully reversible process.
Furthermore, entropy production can be expressed as the sum of classical and quantum contributions arising from changes in population and coherence of the quantum state \cite{santos2019role,mohammadi2024quantum,francica2019role}
\begin{equation}
 \Sigma=\Sigma_q + \Sigma_c
\end{equation}
The classical contribution is written as
\begin{equation}
 \Sigma_c= S(D(\rho^i)||\gamma_S)- S(D(\rho^f)||\gamma_S)
 \label{eq:12}
\end{equation}
and measure how the system approaches the thermal equilibrium.
Indeed, Eq.~(\ref{eq:12}) contain the difference between the initial and final amounts of the athermality monotones, related to those in Eq.~(\ref{eq:6}) with $\alpha=1$, ensuring that the classical entropy production is a non-negative quantity according to the second law of thermodynamics.
The quantum entropy production is expressed as
\begin{equation}
 \Sigma_q= A(\rho^i)-A(\rho^f)
 \label{eq:13}
\end{equation}
and measure how quantum coherence degrades during transport.
$A(\rho)$ is relative entropy of coherence defined as $A(\rho)= S(D(\rho))-S(\rho)$ where $S(\rho)$ denotes the von Neumann entropy, $S(\rho)=-Tr[\rho log(\rho)]$.
Similarly, the non-negativity of  Eq.~(\ref{eq:13}) follows from the fact that relative entropy of coherence is a coherence monotone (Eq.~(\ref{eq:7}) with $\alpha=1$), and introduces an additional second law for quantum coherence.

Furthermore, such a formal splitting of entropy production allows us to highlight the effect of quantum coherence on the dynamics of ion transport.

\section{Results and discussion }
\subsection{\label{sec:level2} Channel-like ion transport }
\begin{figure*} [hbt!]
\includegraphics[width=6in, height=4in]{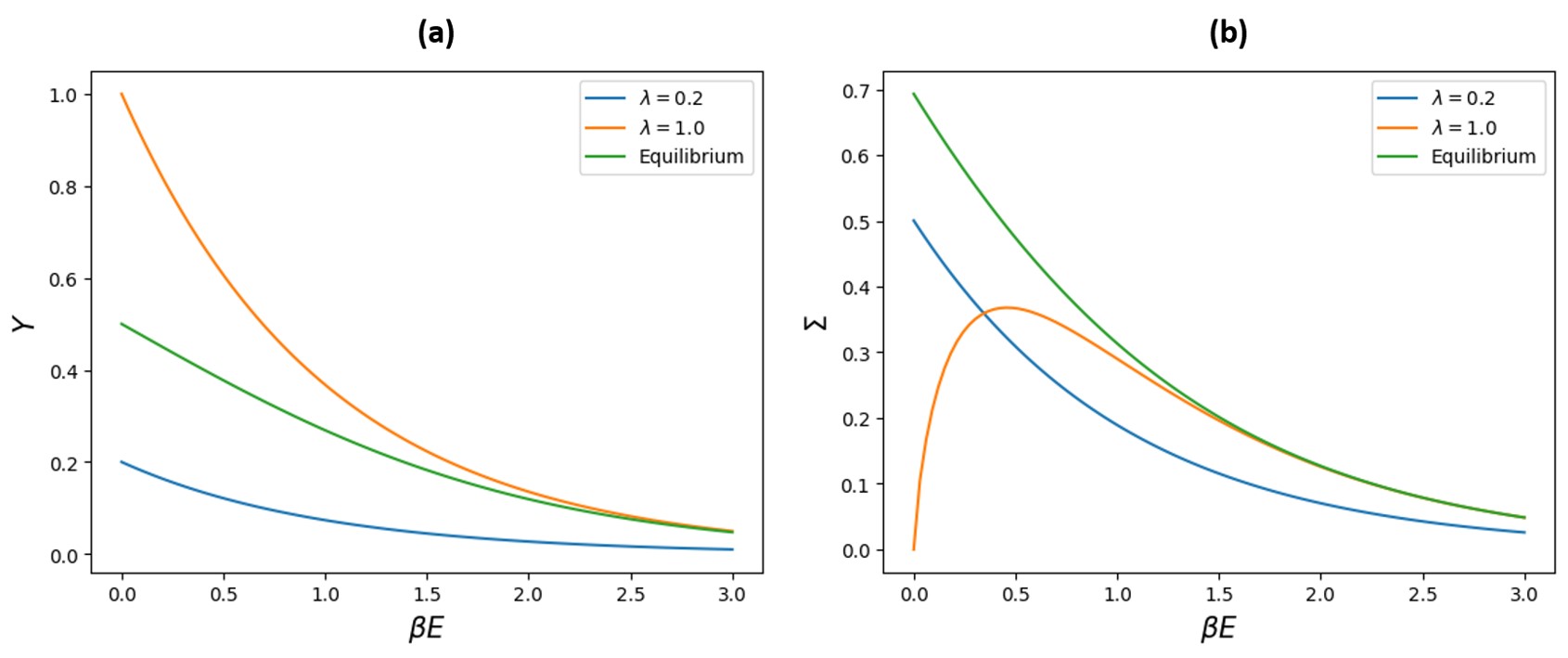}
\caption{\label{fig:2}Comparison between (a) the ion transport yields and (b) the ion transport efficiencies obtained in three different values of parameter $\lambda$ in an incoherent ion channel."Equilibrium" denotes the full thermalization process corresponding to $\lambda=1/Z$.}
\end{figure*}
We will examine the membrane protein functioning as the ion channel here, as depicted in Fig.~\ref{fig:1}(a), and address the distinction between an ion pump and an ion channel in the subsequent subsection.

We assess how the energy gap between two gates $E$, expressed in units of $1/\beta$, affects the yield and efficiency of ion transport in a channel.
In fact, the parameter $E$ provides insight into the energy barrier the cell membrane presents to an ion crossing it.  
This is because ions are more energetically stable on the outside of the membrane than the inside, as they must be dehydrated before entering the membrane interior \cite{gouaux2005principles}.

We also compare results for three different values of parameter $\lambda$. We consider the $\lambda=1$ and $\lambda=0.2$, implying the strong and weak interactions respectively, and $\lambda=1/Z$ that corresponds to full thermalization, bringing the system in the thermal equilibrium distribution.
It was also shown that, for two-level systems, the parameter $\lambda$  provides a simple criteria of Markovianity of thermal operations. Namely, the thermal operation is Markovian when $\lambda \leq 1/Z$ and non-Markovian otherwise \cite{alhambra2019heat,ptaszynski2022non}.
In the following, for all amounts of $\beta E$, $\lambda=0.2$ falls into the Markovian regime, whereas $\lambda=1$ represents the extreme case of the non-Markovian regime, including the most environmental memory effects.

\begin{figure*} [hbt!]
\includegraphics[width=7in, height=4in]{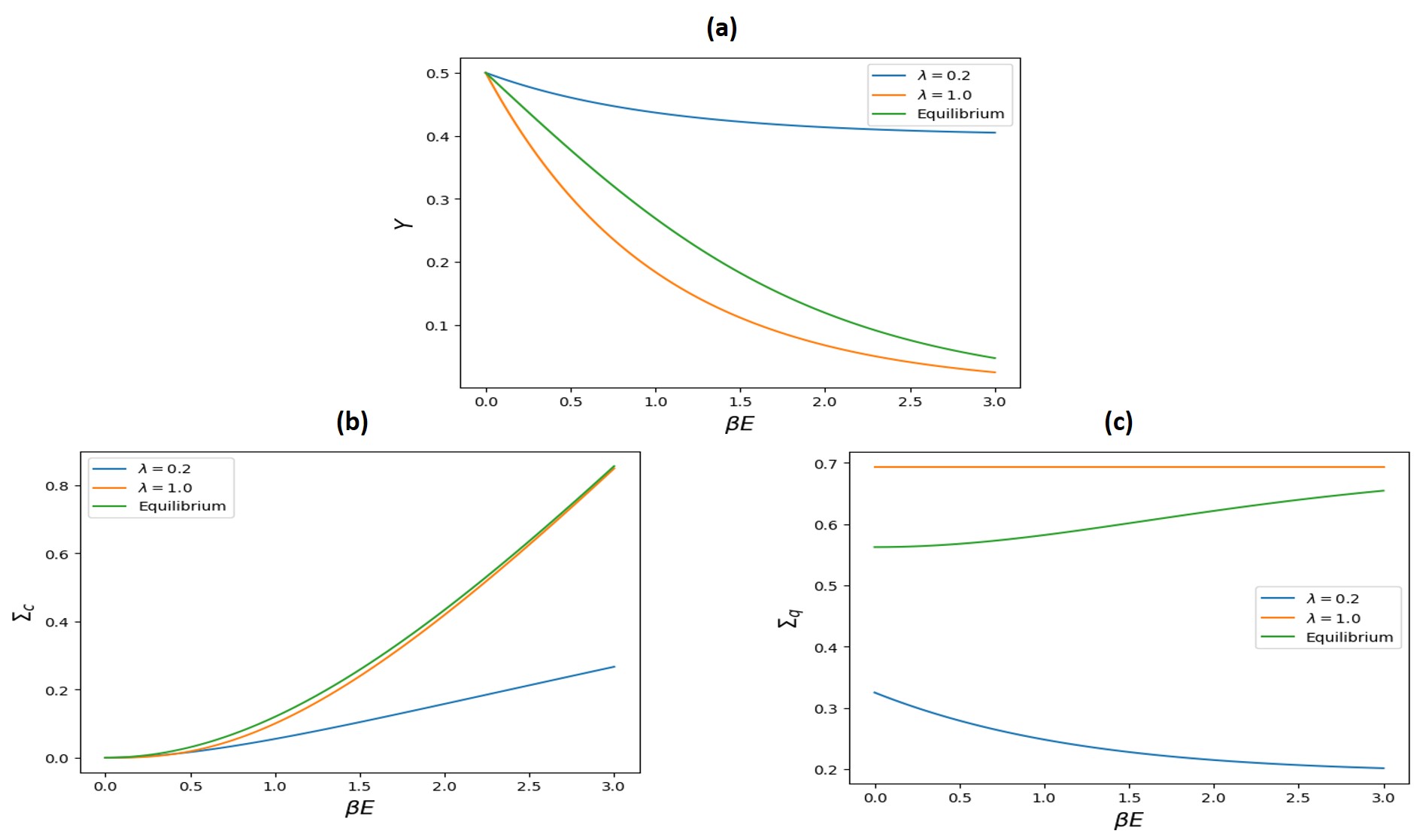}
\caption{\label{fig:3}Comparison between (a) the ion transport yields, (b) the classical and (c) the quantum contributions in the ion transport efficiencies obtained in three different values of parameter $\lambda$ in a coherent ion channel."Equilibrium" denotes the full thermalization process corresponding to $\lambda=1/Z$.}
\end{figure*}

Given an initial pure state of form 
\begin{equation}
 \psi^i= \sqrt{q} |0> + \sqrt{1-q} |1>,    q\in [0,1]
\end{equation}
Figs.~(\ref{fig:2}) and (\ref{fig:3}) show results for two values of $q=1$ and $q=0.5$.

When $q=1$, the ion channel initiates with solely the first gate populated, allowing ions to pass through via a thermal operation that activates the second gate. 
The time translation symmetry inherent in thermal operations ensures that coherence cannot be created during the transport process, resulting in an incoherent ion channel including two separate gates, described by a mixed state.
In this case, the yield values for $\lambda=1$ and $\lambda=0.2$ lie above and below the equilibrium yields, as shown in Fig.~\ref{fig:2}(a). 
In particular, the ion channel can achieve the maximum amount of yield, $Y=1$, supported by maximum environmental assistance, $\lambda=1$, when described by the trivial Hamiltonian where two gates are degenerate, $E=0$. Here, the maximal information encoded in the initial energy eigenstate is the sole thermodynamic resource driving the ion transport dynamics, as no energy change or coherence is present in the system \cite{stratton2023dynamical}.  
However, this scenario is considered rather nonphysical, in the context of ion transport, due to the absence of any energy barrier to ion passage in the channel.
As $E$ increases, the yields decrease monotonically for all values of $\lambda$.

Looking at the Fig.~\ref{fig:2}(b), one can observe a trade-off between the transport yield and the efficiency of the process that produces it. 
A higher value of yield aligns with a higher value of entropy production, indicating a less efficient process, that decrease monotonically as $E$ increases.
An exception, however, is the case when $\lambda=1$ for which the maximum achievable yield at $E=0$ associates with the maximum efficiency, $\Sigma=0$. In this condition, efficiency experiences a temporary decline as $E$ grows by an amount proportional to rise in entropy production from zero. 
At large $E$, however, it coincides with the monotonically decreasing equilibrium efficiency, a situation similarly observed for the transport yields. This illustrates that when encountering a significant energy barrier, an extremely non-Markovian thermal operation cannot outperform thermalizing the system.
Otherwise, without extra quantum resources in an ion channel, non-Markovianity is necessary for achieving the maximum ion transport yield and efficiency in the quantum realm.

When $q=0.5$, the ion channel begins with both gates populated in a maximally coherent state, representing an idealized initial condition for exploring the theoretical implications of coherence in ion transport.
In real biological channels, transient coherence may arise from mechanisms such as electrostatic symmetry at selective binding sites, thermally induced transitions, or external electric field effects \cite{mohseni2014quantum}. However, such coherent states are typically short-lived due to rapid decoherence in biological environments \cite{vaziri2010quantum}.
Comparing Fig.~\ref{fig:3}(a) to Fig.~\ref{fig:2}(a), two differences in yield values obtained in coherent and incoherent ion channels are noticeable.
Firstly, the highest yield in a coherent channel is equivalent to the equilibrium yield, attained at $E=0$, regardless of the specific value assigned to the parameter $\lambda$.
Secondly, for larger values of $E$, the yield values observed in the non-Markovian regime, $\lambda=1$, are lower than the equilibrium yields, while those observed in the Markovian regime, $\lambda=0.2$, exceed the equilibrium values.
The decreasing trend of yields as $E$ increases holds true for all $\lambda$ values in a coherent channel.

A clearer understanding of coherence effects can be gained by examining the efficiency of ion transport within coherent channels.
Figs.~\ref{fig:3}(b) and \ref{fig:3}(c) display the classical and quantum contributions to entropy production in such a process.
It is intuitively expected that quantum coherence will reduce the efficiency of the ion channel, as thermal operations acting on a coherent channel must comply with a more independent positive entropy production constraint, compared to those acting on an incoherent channel, as outlined in Eq.~(\ref{eq:13}). 
More precisely, when there is no energy barrier, $E=0$, the classical contribution does not limit the efficiency of the process, for any values of the parameter $\lambda$. 
However, for the contribution stemming from quantum coherence, channels thermalized with a higher value of $\lambda$ exhibit a lower efficiency.
As the energy barrier rises, classical efficiencies indicate a decline, converging at comparable rates for non-Markovian and full thermalization processes, while the decrease occurs more gradually in the Markovian case.
In contrast, more different behaviors are observed in quantum efficiencies, depending on the various parameter values of $\lambda$. 
For $\lambda=1$, the non-Markovian process consistently exhibits the lowest efficiency, while full thermalization converges towards the non-Markovian efficiency at high values of $E$. On the other hand, when $\lambda=0.2$, the Markovian process shows an increase in efficiency through a reduction in entropy production.
Consequently, the enhanced preservation of quantum coherence within a coherent channel, attributed to non-Markovian thermal operations, is linked to a reduction in the channel's transport efficiency.

These results demonstrate that quantum coherence exerts a detrimental functional influence on ion channels by restricting the maximum achievable yield to the equilibrium value and reducing the efficiency of ion conduction. This is further substantiated by the negative impacts of non-Markovian effects on the transport yield and efficiency of coherent channels, in contrast to positive effects observed in coherent channels.
In fact, in a more technical framework concerning resource theories, non-Markovianity can be viewed as a dynamic resource, representing a valuable aspect of quantum operations rather than quantum states \cite{rivas2014quantum,berk2021resource,lostaglio2017markovian}. This resourcefulness arises here from the ability of non-Markovian operations to preserve information (in incoherent ion channels) and coherence (in coherent ion channels), which are regarded as static resources in the states upon which they act \cite{hsieh2020resource}.
On the other hand, the duration during which the ion channel gates remain simultaneously open is approximately several milliseconds \cite{gadsby2009ion}. Because of the strong interaction of bio-systems with their environment, however, this period is significantly larger than the expected decoherence time during which the gates in an ion channel operate in a superposition state.
All these arguments suggest that non-Markovianity, rather than coherence, is the key quantum resource underlying the high efficiency of ion transport through ion channels.

\subsection{\label{sec:level2} Pump-like ion transport}
We now look at how to differentiate between an ion pump and an ion channel.
To function like a pump, the protein requires the alternate opening of two gates, as depicted in Fig.~\ref{fig:1}(b), preventing them from being populated simultaneously and also from being coherent throughout the transfer process.  
The latter is guaranteed with time translation invariance of thermal operations when the protein begins with only the first gate open, denoted as $\rho^i=|0><0|$.
To establish a fail-safe mechanism, however, the pump must end with only the second gate open, $\rho^f=|1><1|$, thereby inhibiting any channel-like ion flow between the two separate gates.
In other words, the transport yield in a pump should remain unbounded regardless of any changes in model parameters.
Nevertheless, the state transition, $|0><0| \to |1><1|$, is impossible following the second law of thermodynamics, as it would lead to a negative entropy production unless a sufficient amount of energy is provided.
This energy supplement can be included in a battery, conveniently modeled as a two-level system with Hamiltonian $H_W= w|1>_W<1|$, initialised in the exited state $|1>_W<1|$. 
In consequence, the ion transport within the pump can be represented as 

\begin{equation}
 \mathcal{T}(|0><0| \otimes |1>_W<1|)= |1><1| \otimes |0>_W<0|
 \label{eq:15}.
\end{equation}

In fact,  Eq.~(\ref{eq:15}) models the non-spontaneous transport of ions in a pump, which occurs by utilizing energy derived from the hydrolysis of ATP or by being coupled to a spontaneous ion transport occurring in a channel, effectively represented as an incoherent battery system. 

\subsection{\label{sec:level2} Transformation of a pump into a channel}
To model the transformation of a pump into a channel, it is necessary to subvert the fail-safe mechanism of the pump, enabling its two gates to open simultaneously.
This can happen by implementing operations that don’t conserve energy strictly; thereby, coherence can enter the pump's thermodynamics \cite{aaberg2014catalytic,korzekwa2016extraction}.
For this purpose, one needs to utilize an external source of coherence, an ancillary system $R$ with the Hamiltonian $H_R$, and in a state $\sigma_R$ with $[\sigma_R, H_R]\neq 0$.
With the aid of $\sigma_R$, the pump to channel transformation can be realized approximately as an arbitrary unitary operation $U$ that coherently mix the pump's two gates as follows:
\begin{equation}
 Tr_R[V(U) \rho^i_P \otimes \gamma_B \otimes \sigma_R V(U)^\dagger] \approx U \rho_{PB} U^\dagger
\end{equation}
where $V(U)$ is an energy-preserving unitary involving the pump and the thermal bath, and $\rho^i_P=|0><0| \otimes |1>_W<1|$. 
Indeed, the above equation holds with use of ancillary systems with a high degree of coherence in $\sigma_R$,  that can be exemplified as the classical fields like coherent states, i.e $\sigma_R=|\alpha>_R<\alpha|$ that include superpositions over a collection of consecutive eigenstates of $H_R$ \cite{aaberg2014catalytic}.

\section{\label{sec:level1}Conclusion}
In summary, we provided a model for ion transport through the cell membranes, inspired by the inevitable interaction of a quantum system with its environment and the assumption that total energy of the system and the environment is conserved.
This framework allows us to characterize the dynamics of ion transport as a thermal operation that maps the states of the ion-transport protein, which is effectively regarded as a two-level quantum system with each level denoting an open gate that permits ions to either enter or exit the cell.
While this approach simplifies the inherent complexity of biological ion transport, the results presented here are expected to hold as general thermodynamic principles. 
Specifically, we have distinguished between different types of ion-transport proteins: ion channels and ion pumps, with the former encompassing both coherent and incoherent types.
Our results demonstrate that quantum coherence has a detrimental effect on the maximum population of ions that can be transferred, defined as transport yield, and the efficiency of the transport process within an ion channel.
Furthermore, we have shown that non-Markovian effects on ion transport dynamics can enhance both the yield and efficiency of the transport process in an incoherent channel, in contrast to the disadvantage they present in a coherent channel. 
This further supports the idea that coherence may negatively impact the ion channel functionality, while indicating that non-Markovianity, in the absence of additional quantum resources, is the quantum resource that contributes to the high efficiency of ion transport in ion channels.
On the other hand, we have shown that ion transport within a pump constitutes a definitive transition between the protein's ground and excited states, thus ruling out the possibility of quantum coherence in the pump, which is made with the aid of the energy stored in an incoherent two-level battery.
We finally discussed how to model the transformation of a pump into a channel using an external source of coherence. 

This study illustrate that considering quantum properties as the relevant thermodynamic resources for the process of ion transport can elucidate the fundamental difference between ion channels and ion pumps while reinforcing the notion that, despite appearing as completely distinct at first, they actually share many similarities.
This work also suggests the necessity of exploring other biological processes that could benefit from such a resource theory approach to thermodynamics.

\section*{Author Contributions}
\noindent Amin Mohammadi: Conceptualization, Writing – original draft. Afshin Shafiee: Supervision, Writing – review $\&$ editing

\section*{Declaration of interests}
\noindent The authors declare no competing interests.

\section*{REFERENCES}
\bibliography{REFERENCES}
\end{document}